\documentclass[pre,preprint]{revtex4-1}

\usepackage{graphicx}

\begin{document}

\title{Scaling of the dynamics of a homogeneous one-dimensional anisotropic classical Heisenberg model with long-range interactions}

\author{C.~R.~Louren\c{c}o}
\affiliation{Instituto de F\'{i}sica, Universidade de Bras\'\i lia, CP 04455, 70919-970 - Bras\'\i lia, Brazil
and Instituto Federal Bras\'\i lia, Rodovia DF-128, km 21, Zona Rural de Planaltina,
73380-900 - Planaltina, Brazil}
\author{T.~M.~Rocha Filho}
\email{marciano@fis.unb.br}
\affiliation{Instituto de F\'{i}sica and International Center for Condensed Matter Physics, Universidade de Bras\'\i lia, CP 04455, 70919-970 - Bras\'\i lia, Brazil }


\begin{abstract}
The dynamics of quasi-stationary states of long-range interacting systems with $N$ particles can be described by kinetic equations such as the
Balescu-Lenard and Landau equations. In the case of one-dimensional homogeneous systems, two-body contributions vanish as two-body collisions in one dimension
only exchange momentum and thus cannot change the one-particle distribution. Using a Kac factor in the interparticle potential implies a scaling of the dynamics
proportional to $N^\delta$ with $\delta=1$ except for one-dimensional homogeneous  systems. For the latter different values for $\delta$ were reported for a few models.
Recently it was shown by Rocha Filho and collaborators [Phys.\ Rev.\ E {\bf 90}, 032133 (2014)] for the Hamiltonian mean-field model that $\delta=2$ provided that $N$
is sufficiently large, while small $N$ effects lead to $\delta\approx1.7$. More recently Gupta and Mukamel [J.\ Stat.\ Mech.\ P03015 (2011)] introduced a
classical spin model with an anisotropic interaction with a scaling in the dynamics proportional to $N^{1.7}$ for a homogeneous state.
We show here that this model reduces to a one-dimensional Hamiltonian system and that the scaling of the dynamics approaches
$N^2$ with increasing $N$. We also explain from theoretical consideration why usual kinetic theory fails for small $N$ values,
which ultimately is the origin of non-integer exponents in the scaling.
\end{abstract}


\maketitle

\section{Introduction}

Systems with long range interactions may present unusual properties
such as as non-ergodicity, anomalous diffusion, aging, non-Gaussian Quasi-Stationary States (QSS), negative microcanonical
heat capacity, ensemble inequivalence, and more importantly for the present work very long relaxation time to thermodynamic equilibrium of a QSS,
diverging with the number of particles $N$~\cite{newbook,proc1,proc2,proc3,phyr,entropia,ergomarci,transitions1,transitions2}.
A pair interaction potential is long-ranged in $d$ spatial dimensions if it decays at long distances as $r^{-\alpha}$ with $\alpha \leq d$.
The dynamics of such systems can be decomposed in three stages: a violent relaxation into a QSS in a short time,
a slow relaxation of the QSS and the final thermodynamic equilibrium.
In some cases after the violent relaxation the system may also oscillate for a very long or even infinite time around a QSS~\cite{konishi}.
By introducing a Kac factor proportional to $1/N$ in the pair-interaction potential the fluid (Vlasov) limit is well defined and given
by $N\rightarrow\infty$~\cite{kac,braun,balescu,steiner}. The dynamics is exactly described by the Vlasov equation for the one
particle distribution function, while for finite $N$ it is valid only for short times. Collisional terms must be considered for a more accurate description
of the dynamics for finite $N$, leading to kinetic equations such as the Landau or Balescu-Lenard equations~\cite{balescu,lenard,liboff}.

As already noted, the dynamics of relaxation to equilibrium depends on the number of particles in the system and has been extensively studied in the recent
literature~\cite{newbook,phyr,proc1,proc2,proc3,20h,20b,20c,20g,20d,20e,20f,20i,steiner}.
Its dependence on $N$ can be obtained from collisional corrections to the Vlasov equation, i.~e.\ by determining the relevant kinetic equation.
Two-body collisions lead to a collisional integral in the kinetic equations of order $1/N$, and thus relaxation occurs in a
time scale proportional to $N$, an exception being three-dimensional gravity with a relaxation time of order $N/\log N$~\cite{chandrasekhar2,chavanis}.
For one-dimensional homogeneous systems two-body terms in the kinetic equation vanish identically
as collisions between two particles results only in momentum exchange~\cite{eldridge,kadomtsev,sano}.
For instance, the Balescu-Lenard and Boltzmann equations for a homogeneous one-dimensional system with a pair interaction potential
are respectively written as~\cite{liboff}:
\begin{equation}
\frac{\partial}{\partial t}f_1(p_1;t)=\frac{2\pi^2n}{N}\frac{\partial}{\partial p_1}\int {\rm d}p_2\int {\rm d}k
\frac{k^2\tilde V(k)^2}{\left|\varepsilon(k,kp_1) \right|^2}\delta(k(p_1-p_2))
\left(\frac{\partial}{\partial p_1}-\frac{\partial}{\partial p_2}\right)
f_1(p_1;t)f_1(p_2;t),
\label{ballen}
\end{equation}
and
\begin{equation}
\frac{\partial}{\partial t}f_1(p_1;t)=\frac{1}{N}\int dp_2 \left|p_1-p_2\right| \left[f(p_1^\prime;t)f(p_2^\prime;t)-f(p_1;t)f(p_2;t)\right],
\label{boltzeq}
\end{equation}
where $p$ is the one-dimensional momentum variable, $n$ the particle density, $\varepsilon(k,kp_1)$ the dielectric function and
$p_1^\prime$ and $p_2^\prime$ the post-collisional momenta for incoming particles with momenta $p_1$ and $p_2$.
Setting $\varepsilon(k,kp_1)=1$ is equivalent to neglect collective effects and yields the Landau equation.
In both cases the right hand is identically zero due to
the Dirac delta function in the collisional integral in Eq.~(\ref{ballen}), while for the Boltzmann
equation in Eq.~(\ref{boltzeq}) we have $p_1^\prime=p_2$ and $p_2^\prime=p_1$.
In both cases we obtain $\partial f/\partial t=0$ if only two-body collisions are considered,
and the dominant term comes then from three-body or higher order terms.
This has been considered recently for the Hamiltonian Mean Field (HMF) model, resulting in a dynamics
of the homogeneous states scaling with $N^2$~\cite{scaling}, at variance with previous results with scalings proportional
to $N^{1.7}$ and $\exp(N)$ which are due to small $N$ effects~\cite{Zan,nv1,nv2}. The $N^{1.7}$ was also reported for a classical anisotropic Heisenberg model
by Gupta and Mukamel in Ref.~\cite{20d} and the question remains if it is due also to small $N$ effects. In this paper we investigate
this issue for small and large $N$. We re-obtain a $N^2$ scaling for large $N$
as predicted from kinetic theory, while non-integer exponents in the scaling are due to finite $N$ effects, as a  result of the failure of
basic approximations usually considered for the determination of kinetic equation in closed form, as shown below.

The structure of the paper is as follows: in Section~\ref{sec2} we present the model and discuss some of its properties.
The scaling of the dynamics of a QSS is determine numerically in Section~\ref{sec3}.
We address the limits of applicability of kinetic theory in Section~\ref{sec3} and
close the paper with some concluding remarks in Section~\ref{sec5}.

\section{The model}
\label{sec2}

The mean-field classical anisotropic Heisenberg model consists of $N$ classical spins $\vec{S_i} = (S_{ix},S_{ij}, S_{iz})$, $i=1,2,....N$, of unit length globally coupled,
with Hamiltonian~\cite{20d}:
\begin{equation}
H = -\frac{J}{2N}\sum\limits_{i,j=1}^{N}\vec{S_{i}}\cdot \vec{ S_{j}} + D \sum\limits_{i=1}^{N}S_{iz}^2. 
\label{GM}
\end{equation}
The first term in the right-hand side with $J>0$ describes a ferromagnetic mean-field coupling and the second term, a local anisotropy.
Following Gupta and Mukamel we take $J=1$ and $D=15$. Note that the coefficient $1/N$ in the coupling term in Hamiltonian is a Kac factor that makes the energy extensive.
The magnetization of the system is defined by:
\begin{equation}
\vec{m}=\langle\vec{S}\rangle= \frac{1}{N}\sum\limits_{i=1}^{N}\vec{S_{i}}.
\label{maganys}
\end{equation}
Using spherical coordinates the spin components are written as $S_{ix}=\rm{sin}(\theta_i)\rm{cos}(\phi_i)$,
$S_{iy}=\rm{sin}(\theta_i)\rm{sin}(\phi_i)$ and $S_{iz}= \rm{cos}(\theta_i)$,
and the equations of motion are given by:
\begin{equation}
\frac{d \vec{S_i}}{dt} = \{\vec{S_i}, H\},
\end{equation}
with $i=1, 2, ... N$ and the Poisson bracket:
\begin{equation}
\{A,B\} = \sum\limits_{i=1}^{N}\left\{\frac{\partial A}{\partial \phi_i}\frac{\partial B}{\partial S_{iz}},
\frac{\partial A}{\partial S_{iz}}\frac{\partial B}{\partial \phi_i}\right\}.
\end{equation}
Thus
\begin{eqnarray}
\dot{S}_{ix} &=& S_{iy}m_z -  S_{iz}m_y - 2DS_{iy} S_{iz},\nonumber\\
\dot{S}_{iy} &=& S_{iz}m_x -  S_{ix}m_z + 2DS_{ix} S_{iz},\nonumber \\
\dot{S}_{iz} &=& S_{ix}m_y -  S_{iy}m_x.
\label{eqsofmotion}
\end{eqnarray}
These equations of motion admit as first integrals the $z$-component $m_z$ of $\vec{m}$,
the total energy and the the length of each spin. This allows us to rewrite the equations of motion as
\begin{eqnarray}
\dot{\theta}_i & = & m_x \sin(\phi_i)- m_y \cos(\phi_i), \nonumber\\
\dot{\phi}_i  & = & m_x \cot(\theta_i)\cos(\phi_i)+ m_y \cot(\theta_i)\sin(\phi_i)-m_z+2D\cos(\theta_i).
\label{eqmotion}
\end{eqnarray}
As a first result we observe that these equations are canonical and derive from the Hamiltonian:
\begin{eqnarray}
H & = &-\sum_{i=1}^N\left[m_x\cos(\phi_i)\sqrt{1-S^2_{iz}}+m_y\sin(\phi_i)\sqrt{1-S^2_{iz}}\right.\nonumber \\
& & \left.+m_zS_{iz} - DS^2_{iz}\right].
\label{hamiltoniana}
\end{eqnarray}
were $p_i\equiv\cos \theta_i=S_{iz}$ and $q_i\equiv\phi_i$ are canonically conjugate and correspond to the momentum and position variables respectively.
As a consequence the model is effectively one-dimensional and thus explains why a scaling proportional to $N^\delta$ with $\delta\neq 1$ is observed.
As the model is effectively one-dimensional and Hamiltonian the tools of kinetic theory can be used to derive a kinetic equation,
as described for instance in Ref.~\cite{balescu}. The first consequence of this fact is that for a homogeneous state in $\phi$,
the collisional integral proportion to $1/N$ of the Balescu-Lenard equation vanishes, and one must go to the next order in an expansion in
powers of $1/N$ (see Ref.~\cite{scaling} and references therein).

\section{Scaling of the dynamics}
\label{sec3}

In order to study the dynamics of a homogeneous state, and for comparison purposes, we use here the same initial condition as in Ref.~\cite{20d},
a  waterbag state (uniform distribution) in the intervals $\phi\in[0,2\pi)$
and $\theta\in[\pi/2-a,\pi/2+a]$, with energy per particle:
\begin{equation}
e=\frac{D}{3}\sin^2 a,
\label{energypp}
\end{equation}
and $a$ chosen such that $e=0.24$. The state is spatially homogeneous and stable for this energy.
From Ref.~\cite{scaling} the expected scaling of the dynamics of this QSS is $N^2$. On the other hand
Gupta and Mukamel obtained from numerical simulations a different scaling in $N^{1.7}$. We argue that, similar to what occurs in the HMF model,
the $N^{1.7}$ scaling only occurs for sufficiently small number of particles, while for larger $N$ the scaling becomes
proportional to $N^2$.

\begin{figure}[ht]
\includegraphics[width = 10cm]{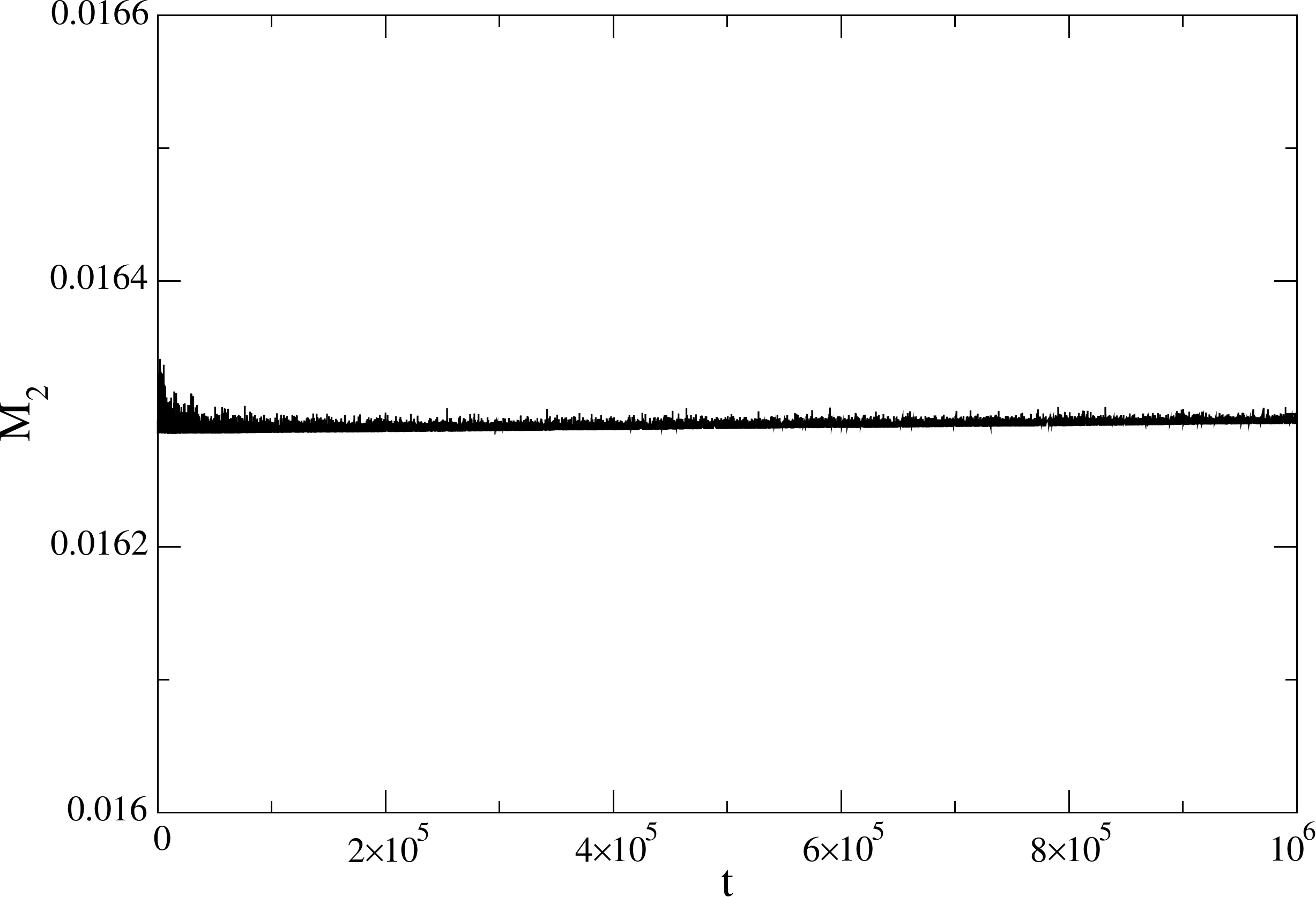}
\caption{Second statistical moment $M_2$ of variable $\theta$ for $N=100\,000$.}
\label{fig0}
\end{figure}

In a homogeneous stable state the spatial distribution for variable $\phi$ is always
uniform up to small fluctuations, but the distribution for variable $\theta$ slowly varies with time towards thermodynamic equilibrium~\cite{scaling}.
As a consequence, the dynamics can be probed by the statistical moments $M_n=\langle(\theta-\langle\theta\rangle)^n\rangle$. Odd moments of $\theta$ vanish
for an even distribution in $\theta$ as is the case here.
Figure~\ref{fig0} shows the second moment $M_2$ as a function of time. It varies very slowly for the states considered here
(it is almost a constant of motion) so we consider the time evolution of the fourth moment $M_4$ which is more responsive to small changes in the
statistical state of the system. In Ref.~\cite{20d} Gupta and Mukamel considered the average $\langle \cos\theta\rangle$ which is
more difficult to characterize the small variations in the distribution function of $\theta$ (compare for instance Fig.~3 of their paper to
our Figures~\ref{fig0} and~\ref{fig1} below).
The equations of motion in Eq.~(\ref{eqmotion}) are solved using a parallel implementation of a fourth order Runge-Kutta algorithm in a graphics processing unit
using the CUDA extension to the C language~\cite{cuda,marcicomp}. This allows us to perform simulations with a much greater number of particles than
considered in Ref.~\cite{20d}. The time-step used is $\delta t=0.01$ and ensures a maximum relative error in the energy or order $10^{-4}$.
Figure~\ref{fig1} shows the time evolution of $M_4$ for different number of particles up to $N=100\,000$. Figures~\ref{fig2} and~\ref{fig3} show
the same results but with $1/N^{1.7}$ and $1/N^2$ time rescalings, respectively. A better data collapse is obtained for the $N^2$ scaling.

\begin{figure}[ht]
\includegraphics[width = 10cm]{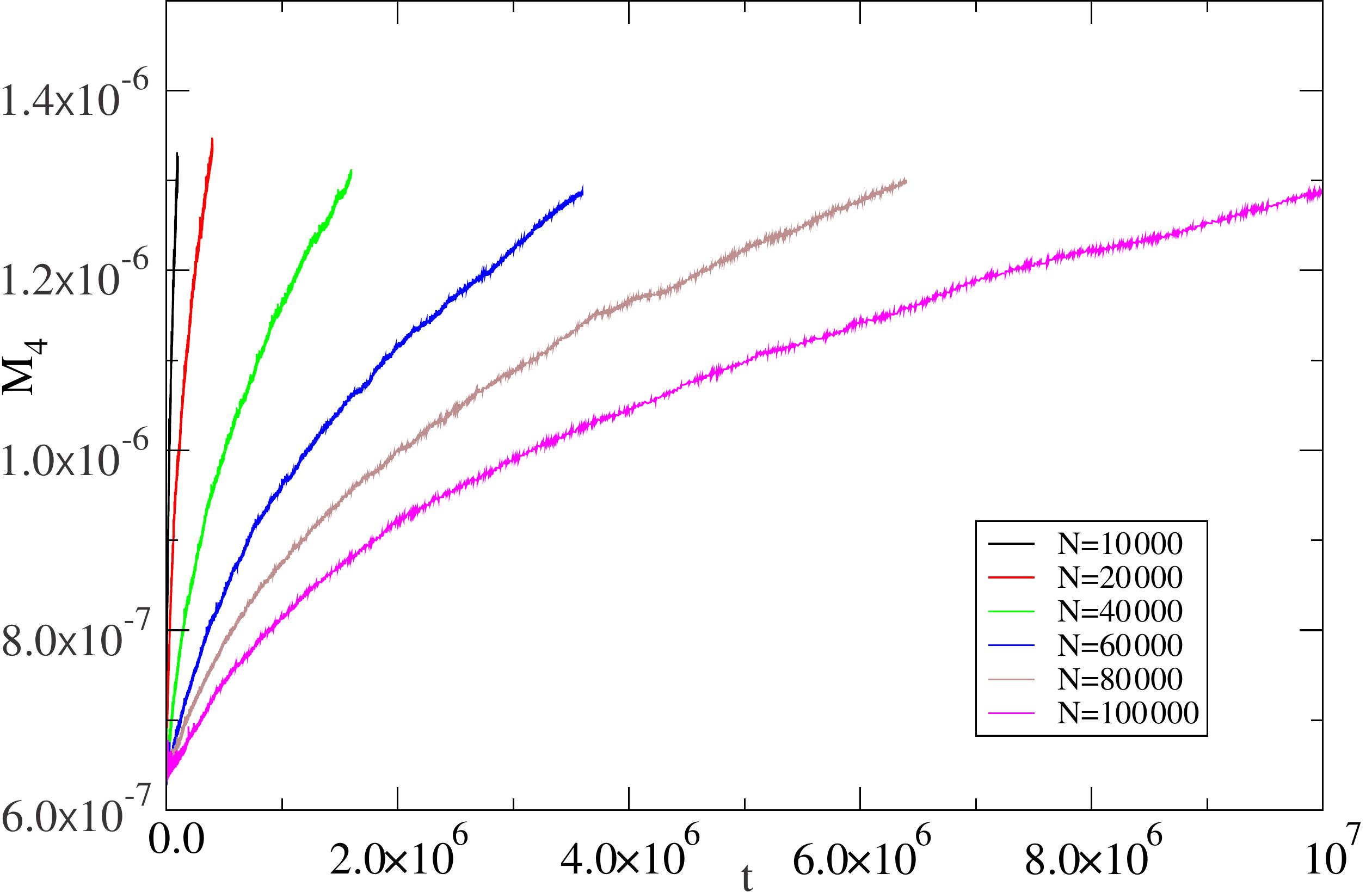}
\caption{(Color online) Moment $\langle M_4\rangle$ of variable $\theta_i$ as a function of time for different numbers of particles
$N=10\,000,\ 20\,000,\ 40\,000,\ 60\,000,\ 80\,000,\ 100\,000$.}
\label{fig1} 
\end{figure}

\begin{figure}[ht]
\includegraphics[width = 10cm]{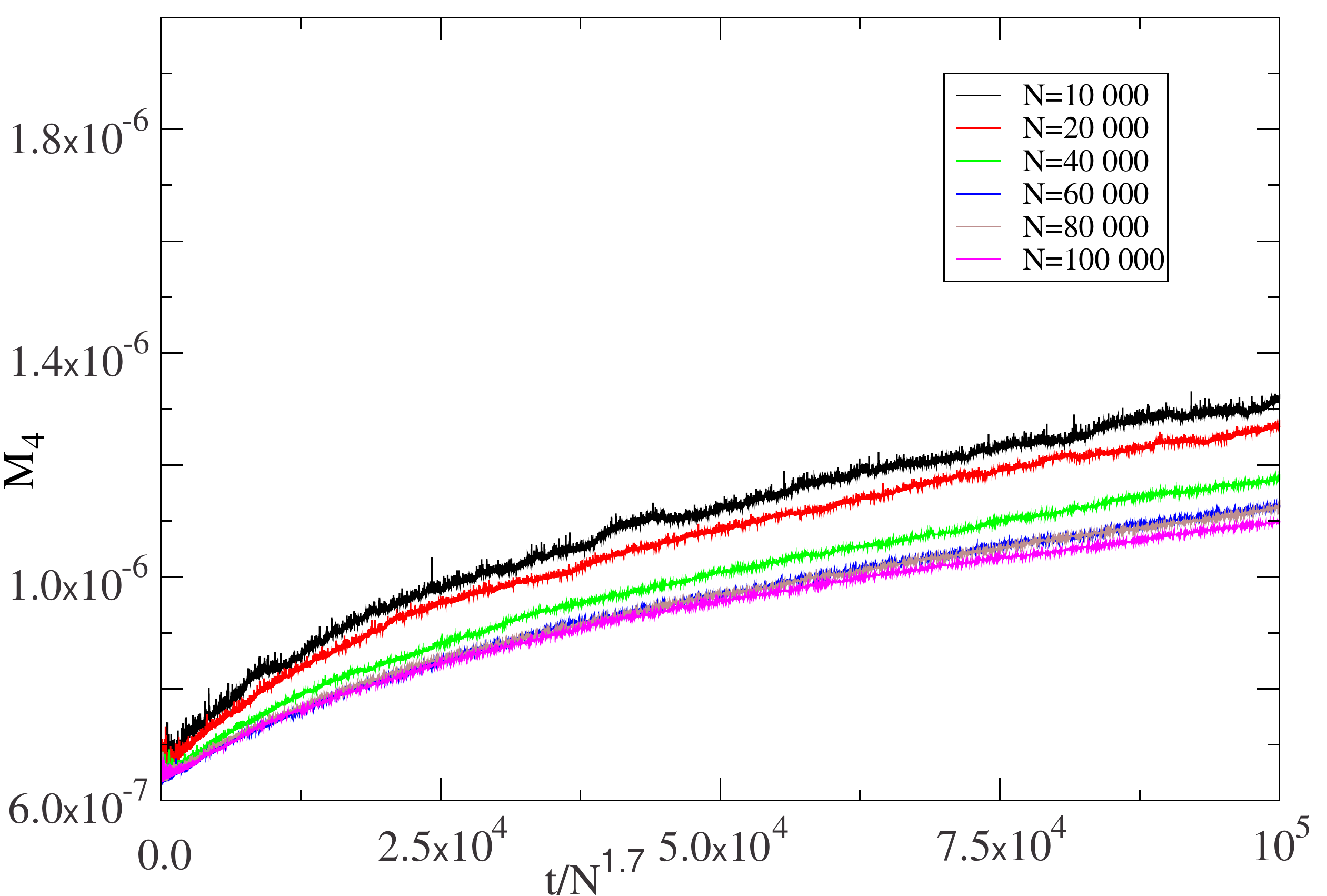}
\caption{(Color online) Same as in Fig.~\ref{fig1} but with a time rescaled by $(N-10\,000)^{-1,7}$.}
\label{fig2}
\end{figure}

\begin{figure}[ht]
\includegraphics[width = 10cm]{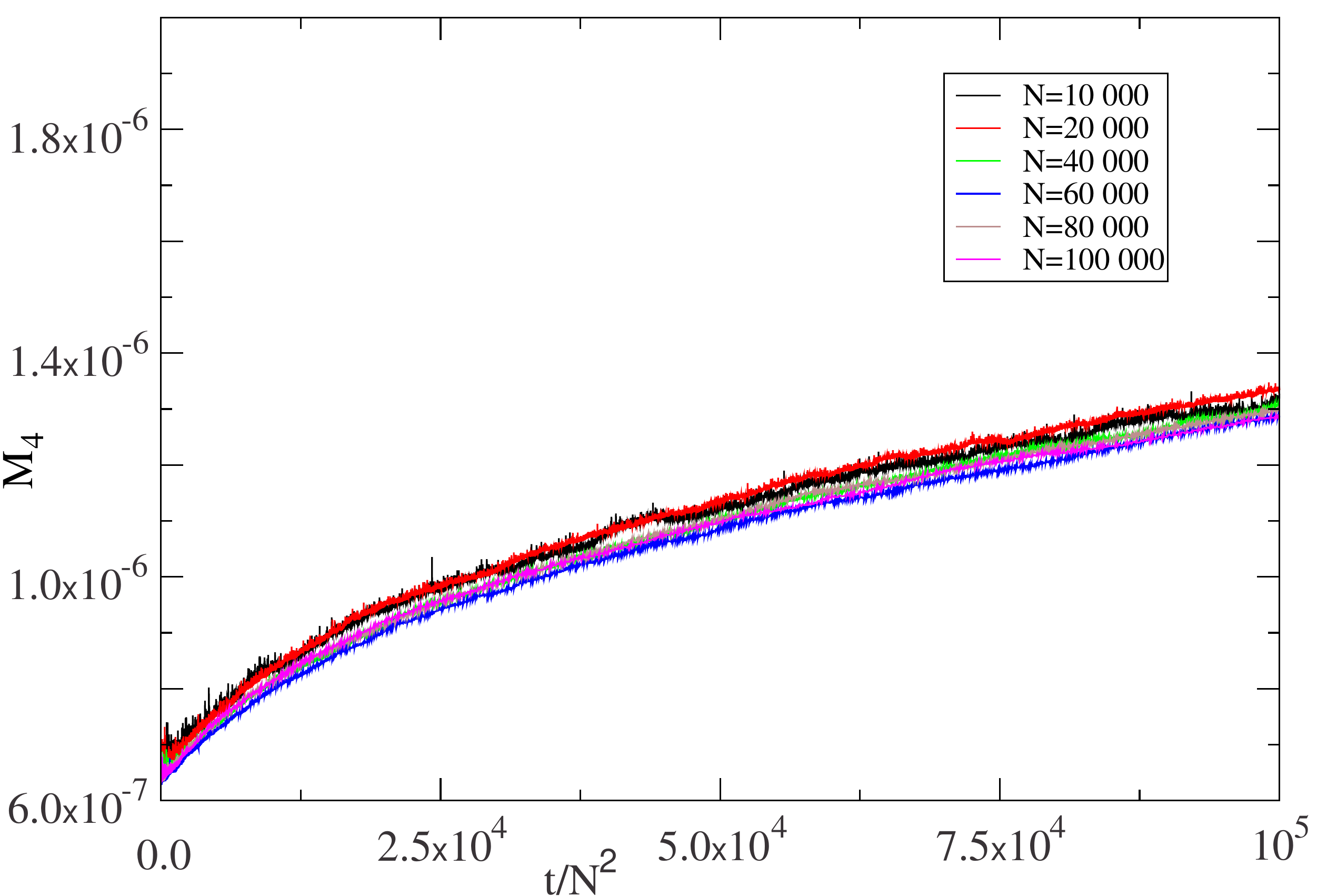}
\caption{(Color online) Same as in Fig.~\ref{fig1} but with a time rescaled by $(N-10\,000)^{-2}$.}
\label{fig3}
\end{figure}

\begin{figure}[ht]
\includegraphics[width = 10cm]{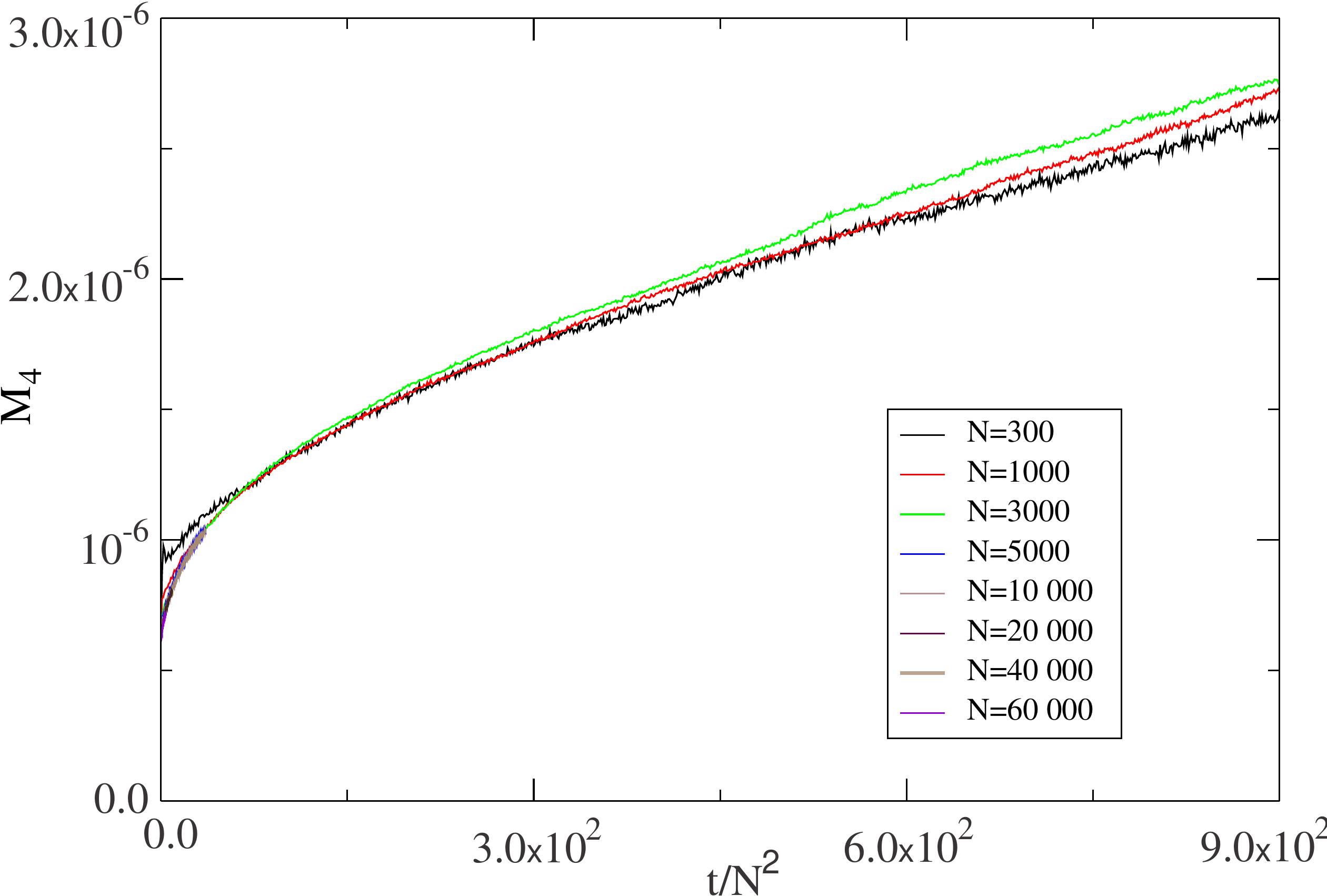}
\caption{(Color online) Moment $M_4$ of the $\theta_i$ as a function of time for $N=300,\ 1000,\ 3000,\ 5000,\ 10\,000,\ 20\,000,\ 40\,0000, \ 60\,000$
with a time rescaling $(N-300)^{-2}$. The number of realization varies from $300$ for $N=300$ to $25$ for $N=60\,000$.}
\label{fig4}
\end{figure}

In order to compare quantitatively ours with previous results, we performed a series of simulations with the same number of particles as in Ref.~\cite{20d} but also considering
values of $N$ up to $60\,000$. By averaging over many realizations we compare the time evolution of $M_4$
for a given value of $N$ with the previous smaller number of $N$ in the simulations, and perform a least squares fit
for the difference between both time series rescaled by $1/N^\delta$.
The results are shown in Table~\ref{tabela} and corroborate, up to some small deviations, a scaling in $N^2$.
Figure~\ref{fig4} shows the statistical moment $M_4$ for the same number of particles as in Table~\ref{tabela}
with time scaled as $1/N^2$ with a very good data collapse for $N\geq 5000$.

We note that, in Ref.~\cite{20d}, Gupta and Mukamel determined the scaling behavior considering the values $N=300, 1000, 3000, 5000$. The difference of theirs and our results for the case $N=3000$ and $5000$ comes from the fact that considering the magnetization as a relevant variable yields more imprecise results then when considering the statistical moments of the momenta variables (see also the discussion of this in Ref.~\cite{marcicomp}).

\begin{table}[ht]
\centering
\vspace{0.2cm}
\begin{tabular}{|c|c|c|}
$N_1$ & $N_2$ & $\delta$  \\
\hline  \hline
300 & 1000 &  1.767       \\
1000 & 3000 & 1.797   \\
3000 & 5000 & 2.015           \\
5000 & 10\,000 & 2.056        \\
10\,000 & 20\,000 & 2.072  \\
20\,000 & 40\,000 & 2.066     \\
40\,000 & 60\,000 & 2.096
\end{tabular}
\caption{Best scaling in $N^\delta$ for the moment $M_4$ between a pair of simulated data with $N_1$ and $N_2$ particles.}
\label{tabela}
\end{table}

\section{Limitations of kinetic theory}
\label{sec4}

Our results are in agreement with what is expected from a kinetic theory derived from the BBGKY hierarchy in a series expansion in power of $1/N$.
Two and three particle correlation functions contribute with terms proportional to $1/N$ and $1/N^2$, respectively.
As two-particle contributions to the kinetic equation vanishing in the present case
one must retain the contributions from three-particle collisions which are proportional to $1/N^2$.
These considerations are based on the introduction of the Kac factor in the Hamiltonian and
the scaling proportional to $N^{-1.7}$ reported by Gupta and Mukamel is re-obtained here for smaller values of $N$.
This unusual scaling stems on the failure for small $N$ of the Markovianization hypothesis used in the determination of
the Balescu-Lenard and Landau equations, which requires that the force auto-correlation function (for homogeneous systems) differs significantly
from zero only for very short times if compared to the dynamical time scale over which the one-particle distribution function varies significantly.
Let us show this explicitly for the simpler case of the Landau equation, i.~e.\ for weak coupling, as the same kind of approximations are
used in the deduction of the Balescu-Lenard equation (see Ref.~\cite{balescu} for a thorough discussion on these assumptions).

The $N$-particle distribution function $f_N({\bf r}_1,{\bf v}_1,\ldots,{\bf r}_N,{\bf v}_N;t)$ is the probability density in the
$N$-particle phase space for a particle at time $t$ to have position ${\bf r}_i$ and momentum ${\bf p}_i$.
Defining the $s$-particle distribution function by
\begin{equation}
f_s\equiv f_s({\bf r}_1,{\bf v}_1,\ldots,{\bf r}_s,{\bf v}_s;t)=\int d{\bf r}_{s+1}d {\bf v}_{s+1}\cdots d{\bf r}_Nd{\bf v}_N\:
f_N({\bf r}_1,{\bf v}_1,\ldots,{\bf r}_N,{\bf v}_N;t).
\label{deffs}
\end{equation}
where ${\bf r}_i$ and ${\bf p}_i$ are the position and momentum vectors of particle $i$ in $d$ dimensions. Liouville equation then implies
that the reduced distribution functions satisfy the BBGKY hierarchy~\cite{liboff,balescu}:
\begin{equation}
\frac{\partial}{\partial t} f_s=\sum_{j=1}^s\hat{K}_j f_s+\frac{1}{N}\sum_{j<k=1}^s\hat\Theta_{jk}f_s
+\frac{N-s}{N}\sum_{j=1}^s\int d {\bf r}_{s+1}d {\bf v}_{s+1}\hat\Theta_{j,s+1}f_{s+1},
\label{bbgky2}
\end{equation}
where
\begin{equation}
\hat{K}_j=-{\bf v}_j\cdot\nabla_j,\hspace{5mm}\hat\Theta_{jk}=-\nabla_jV({\bf r}_j-{\bf r}_k)\partial_{jk},
\hspace{5mm}
\partial_{jk}\equiv\frac{\partial}{\partial {\bf v}_j}-\frac{\partial}{\partial {\bf v}_k},
\label{defcalkth}
\end{equation}
and $\nabla_j$ is the gradient operator for the position of particle $j$.
In order to obtain a closed kinetic equation for the one-particle distribution function $f_1$ we have to determine the functional dependence
of $f_2$ on $f_1$ (Bogolyubov hypothesis~\cite{liboff}). This can be accomplished in the present framework by writing the reduced distribution functions
in the form of a cluster expansion, which for a homogeneous system is given by:
\begin{eqnarray}
 f_2({\bf v}_1,{\bf v}_2,{\bf r}_1-{\bf r}_2) & = & f_1({\bf v}_1)f_1({\bf v}_2)+C_2({\bf v}_1,{\bf v}_2,{\bf r}_1-{\bf r}_2),
\label{cluster2}
\\
 f_3({\bf v}_1,{\bf v}_2,{\bf v}_3,{\bf r}_1-{\bf r}_2,{\bf r}_2-{\bf r}_3) & = & f_1({\bf v}_1)f_1({\bf v}_2)f_1({\bf v}_3)+
\sum_{P(1,2,3)}f_1({\bf v}_1)C_2({\bf v}_2,{\bf v}_3,{\bf r}_2-{\bf r}_3)
\nonumber\\
 & & +C_3({\bf v}_1,{\bf v}_2,{\bf v}_3,{\bf r}_1-{\bf r}_2,{\bf r}_2-{\bf r}_3),
\label{cluster3}
\end{eqnarray}
and so on, where the time dependence is kept implicit,  $P(1,2,3)$ stands for permutations of particles $1$, $2$ and $3$ and $C_s$ is the $s$-particle correlation function.
Let us consider a parameter $\lambda\ll1$ characterizing the strength of the interaction, i.~e.\ $V={\cal O}(\lambda)$.
A two-particle correlation  requires the interaction of two particles to be created and therefore $C_2$ is of order $\lambda$.
A three-particle correlation requires the interaction between two pairs of particles and thus $C_3$ is of order $\lambda^2$, and so on.
By considering the case $s=1$ in Eq.~(\ref{bbgky2}) and using Eq.~(\ref{cluster2}) we have:
\begin{equation}
\frac{\partial }{\partial t}f_1({\bf v}_1;t)=\frac{N-1}{N}\int d{\bf v}_2d{\bf r_2}\:\hat\Theta_{12}
\left[f_1({\bf v}_1;t)f_1({\bf v}_2;t)+C_2({\bf v}_1,{\bf v}_2,{\bf r}_1-{\bf r}_2;t)\right].
\label{landau1}
\end{equation}
The two-particle correlation function is the solution of the equation obtained by replacing Eq.~(\ref{cluster3})
into Eq.~(\ref{bbgky2}) for $s=2$ and discarding higher order terms containing three-particle correlations:
\begin{equation}
\left(\frac{\partial }{\partial t}-\hat K_1-\hat K_2\right)C_2({\bf v}_1,{\bf v}_2,{\bf r}_1-{\bf r}_2;t)=\hat\Theta_{12}f_1({\bf v}_1;t)f_1({\bf v}_2;t).
\label{landau2}
\end{equation}
Its solution can be written as:
\begin{eqnarray}
\lefteqn{\hspace{-10mm}C_2({\bf v}_1,{\bf v}_2,{\bf r}_1-{\bf r}_2;t)=e^{\left(\hat K_1+\hat K_2\right)t}C_2({\bf v}_1,{\bf v}_2,{\bf r}_1-{\bf r}_2;0)}
\nonumber\\
 & & +\int_0^t dt\:e^{\left(\hat K_1+\hat K_2\right)\tau}\hat \Theta_{12}f_1({\bf v}_1;t-\tau)f_1({\bf v}_2;t-\tau).
\label{landau3}
\end{eqnarray}
The first term in the right-hand side of Eq.~(\ref{landau3}) is a transient term due to correlation at $t=0$ and dies out rapidly~\cite{balescu}.
By replacing Eq.~(\ref{landau3}) into Eq.~(\ref{landau1}) and noting that the mean-field force vanishes in a homogeneous state, we obtain (using $N-s\rightarrow N$ for large $N$):
\begin{eqnarray}
\frac{\partial }{\partial t}f_1({\bf v}_1;t) & = & \int_0^t id\tau\int d{\bf v}_2d{\bf r_2}\:\hat\Theta_{12}
e^{\left(\hat K_1+\hat K_2\right)\tau}\hat \Theta_{12}f_1({\bf v}_1;t-\tau)f_1({\bf v}_2;t-\tau)
\nonumber\\
 & = & \int_0^t d\tau\int d{\bf v}_2d{\bf r_2}\:\partial_{12}\nabla_1V({\bf r}_{12})
e^{\left(\hat K_1+\hat K_2\right)\tau}\nabla_1V({\bf r}_{12})\partial_{12}f_1({\bf v}_1;t-\tau)f_1({\bf v}_2;t-\tau)
\nonumber\\
 & = & \int_0^t d\tau\int d{\bf v}_2d{\bf r_2}\:\partial_{12}\nabla_1V({\bf r}_{12})
\nabla_1V({\bf r}_{12}-{\bf v}_{12}\tau)\partial_{12}f_1({\bf v}_1;t-\tau)f_1({\bf v}_2;t-\tau)
\label{landau4}
\end{eqnarray}
with ${\bf r}_{12}\equiv{\bf r}_1-{\bf r}_2$ and ${\bf v}_{12}\equiv{\bf v}_1-{\bf v}_2$.
The force auto-correlation of the ${\bf F}({\bf r},t)$ at position ${\bf r}$ is defined by
\begin{equation}
{\cal C}(t)\equiv\langle F(t)F(0)\rangle=\int d{\bf r}\:{\bf F}({\bf r},0){\bf F}({\bf r},t)=\int d{\bf r}\:\nabla V({\bf r}-{\bf v}_{12}t)\:\nabla V({\bf r}).
\label{landau5}
\end{equation}
Thence we have:
\begin{equation}
\frac{\partial }{\partial t}f_1({\bf v}_1;t)=\int_0^t d\tau\int d{\bf v}_2\:\partial_{12}\langle F(\tau)F(0)\rangle\:
\partial_{12}f_1({\bf v}_1;t-\tau)f_1({\bf v}_2;t-\tau).
\label{landau6}
\end{equation}
This is a master equation which is non-Markovian as it depends on $f_1$ at previous times form $0$ to $t$.
To obtain a true (Markovian) kinetic equation the usual procedure is to assume
that the dynamic time scale $t_d$ over which the one-particle distribution function $f_1$ varies significantly is much greater than the
time scale $t_c$ such that the force auto-correlation is sufficiently small. In this case, one can replace $f_1({\bf v};t-\tau)$ in the integrand
in Eq.~(\ref{landau6}) by $f_1({\bf v}_1;t)$, which corresponds to the ballistic approximation (free motion for a homogeneous system),
and extend the time integration to infinity. We then finally obtain the Landau equation:
\begin{equation}
\frac{\partial }{\partial t}f_1({\bf v}_1;t)=\int_0^\infty d\tau\int d{\bf v}_2\:\partial_{12}\langle F(\tau)F(0)\rangle\:
\partial_{12}f_1({\bf v}_1;t)f_1({\bf v}_2;t).
\label{landafin}
\end{equation}
This form will suffice for the present discussion.
The same type of considerations are also necessary in the determination of the Balescu-Lenard equation~\cite{balescu,lenard}.
As discussed above, for a one-dimensional homogeneous system these corrections
vanish and one must go one order further in the $1/N$ expansion.
Usually one always considers a Markovianization procedure taking into account the time scales such that $t_d\gg t_c$.
A failure of this condition implies, among other things, that the collisional integral does not
vanish exactly for a one-dimensional homogeneous system, and one should expect that the scaling of the dynamics is therefore affected.

\begin{figure}[ht]
\includegraphics[width = 16.5cm]{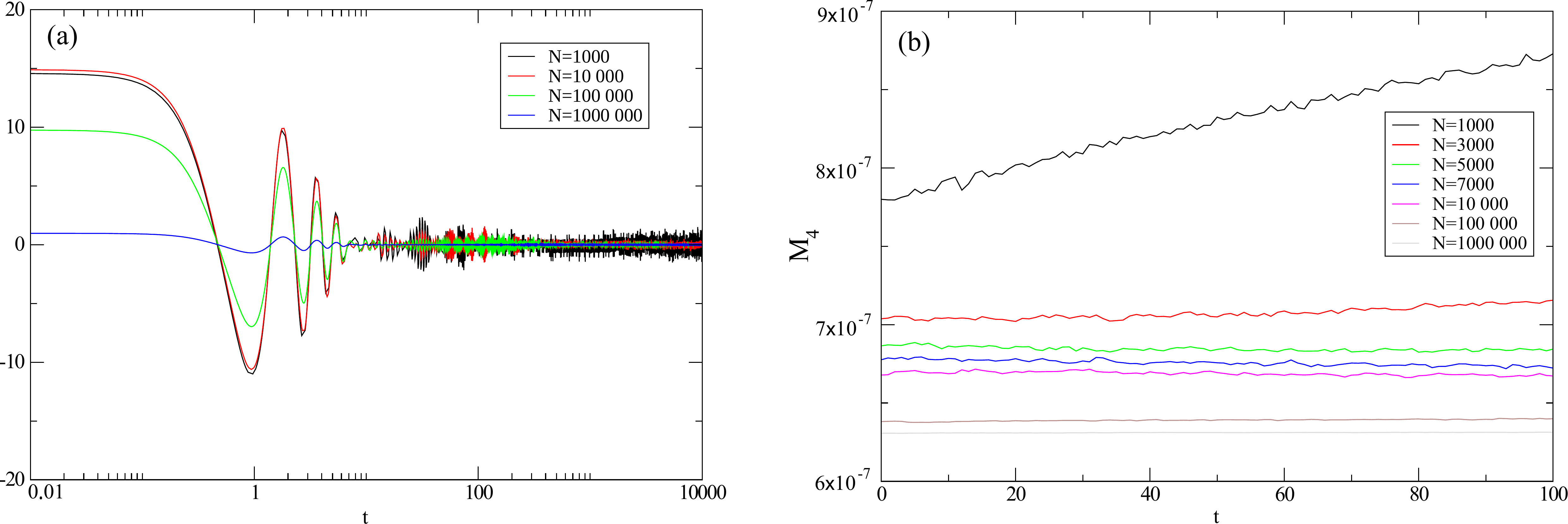}
\caption{(Color online) a) Force auto-correlation ${\cal C}(t)$ as a function of time for different values of $N$. b) Time evolution for
the fourth moment $\langle M_4\rangle$ of variable $\theta$ averaged over $1000$ realizations except for $N=100\:000$ and $N=1\:000\:000$ with $300$ and $200$ realizations,
respectively. The initial conditions are the same homogeneous state as in Fig.~\ref{fig1} thermalized up to $t=20.0$ before starting the simulations shown here.}
\label{fig5}
\end{figure}

In order do address this point, we compute the force auto-correlation function from numeric simulations by:
\begin{equation}
{\cal C}(t)=\frac{1}{N}\sum_{i=1}^N F_i(t)F_i(0),
\label{forceautodef}
\end{equation}
where $F_i(t)$ is the force on particle $i$ at time $t$ due to all other particles. Figure~\ref{fig5}a shows ${\cal C}(t)$
for different number of particles for the present model and the time evolution of $\langle M_4\rangle$ for variable $\theta$.
We observe that the time required for a significant decrease of ${\cal C}(t)$, i.~e.\ the correlation time $t_c$, is roughly the same for all values of $N$, while
the dynamical time $t_d$ is smaller for smaller $N$ as shown in the right panel of Fig.~\ref{fig5}.
In this way, the correlation time can become of the same order of magnitude as the dynamical time, breaking down the Markovian condition, and therefore the usual
derivation of Kinetic equations from the BBGKY hierarchy is no longer valid. Figure~\ref{fig5}b show the fourth moment $M_4$ of variable $\theta$ and it becomes
evident that Markovianity is not valid for $N=1000$ and $N=3000$, while it is approximately valid for $N=5000$. For $N\geq10\:000$ the system is clearly Markovian,
in agreement with the results in Table~\ref{tabela}. This explains why a different scaling in $N^\delta$ of the dynamics with $\delta\neq2$
is observed for homogeneous one-dimensional systems for small $N$~\cite{scaling}.

\section{Concluding remarks}
\label{sec5}

We shown in this paper that the mean-field anisotropic Heisenberg model introduced by Gupta and Mukamel in Ref.~\cite{20d} is effectively
a one-dimensional classical Hamiltonian system, and the dynamics of a QSS scales as $N^2$ for large $N$ while the scaling in $N^{1.7}$ previously
reported is due to small $N$ non-Markovian effects in the dynamics. For large $N$ a kinetic
equation for a homogeneous one-dimensional long-range interacting system must consider three-body collisions, which are of order $1/N^2$.
This approach is only valid if $N$ is sufficiently large such that the contribution of strictly two-body collisions
does vanish, while for small $N$ the arguments leading to the $N^2$ scaling fail.
The small $N$ case can be tackled using an approach developed by Ettoumi and Firpo by determining the diffusion coefficient in terms of action variables
who used a mean passage time approach~\cite{ettoumi} and obtained a $N^{1.7}$ scaling for the Hamiltonian mean Field model~\cite{hmforig}.
Based on time evolution of the auto-correlation of the force for the homogeneous case, one can consider if a similar behavior occurs
for non-homogeneous one-dimensional and for higher dimensional systems, and whether and how it influences the scaling  for small $N$.
This will be the subject of a separate publication.
This also raises the question whether similar effects might have a role in astrophysics.
Indeed smaller globular clusters can have a number of stars of the order of just a few thousand, as opposed to $10^{10}$
stars in a typical galaxy. Other long-range interacting systems may also have a similar behavior.
More recently, Gupta and Mukamel introduced a different model of classical spins in a sphere described bu a two-dimensional Hamiltonian~\cite{gm2}
and also displaying a scaling of the dynamics of a homogeneous QSS in $N^{1.7}$. Taking into account the discussion in the present paper and in
Ref.~\cite{scaling} this is a strong indication that this model is in fact effectively one-dimensional, which is still to be shown.

\section{Acknowledgments}
The authors acknowledge partial financing by CAPES (Brazilian government agency). TMRF would like to thank B.~Marcos for fruitful discussions.

\end{document}